\documentclass[prl,twocolumn,showpacs]{revtex4}

\usepackage{graphicx}
\begin{document}

\title{
   Transition from Abelian to Non-Abelian Quantum Liquids
   in the Second Landau Level}

\author{
   Arkadiusz W\'ojs}

\affiliation{
       TCM Group, Cavendish Laboratory, University of Cambridge, 
       Cambridge CB3 0HE, United Kingdom, and\\
       \mbox{Institute of Physics, Wroclaw University of Technology,
       Wyb.\ Wyspia\'nskiego 27, 50-370 Wroclaw, Poland}}

\begin{abstract}
The search for non-Abelian quantum Hall states in the second Landau 
level is narrowed to the range of filling factors $7/3<\nu_e<8/3$.
In this range, the analysis of energy spectra and correlation functions, 
calculated including finite width and Landau level mixing, supports the 
prominent non-Abelian candidates at $\nu_e=5/2$ (Moore--Read) and $12/5$
(Read--Rezayi).
Outside of it, the four-flux noninteracting composite fermion model is
validated.
The borderline $\nu_e=7/3$ state is adiabatically connected to the 
Laughlin liquid, but its short-range correlations and charge 
excitations are different.
\end{abstract}
\maketitle

The incompressible quantum liquids (IQLs) \cite{Laughlin83} formed 
in a high magnetic field $B$ by two-dimensional electrons filling 
various fractions $\nu$ of different Landau levels (LL$_n$, $n=0$, 
1, \dots) have been the subject of extensive studies ever since 
the famous discovery of the fractional quantum Hall (FQH) effect 
\cite{Tsui82}.
The most recent storm of interest is motivated by the concept of 
``topological quantum computation'' \cite{Kitaev03,Nayak08}
employing non-Abelian statistics of the Moore--Read ``pfaffian'' 
wave function \cite{Moore91,Tserkovnyak03} believed to describe 
the FQH state in a half-filled LL$_1$.
Other wave functions with different complexities of braiding statistics 
have also been proposed \cite{Read99,Moller08a,Bonderson08}, elevating
the convincing demonstration of non-Abelian statistics in a real 
physical system to the challenge of greatest current importance.

With the partially filled LL$_0$ successfully described by the 
composite fermion (CF) theory \cite{Jain89} and with ordered electron 
phases favored in higher LLs, signatures of non-Abelian statistics 
are most strongly anticipated in the few known IQLs in LL$_1$.
Crucial recent experiments in this LL include confirmation of the 
quasiparticle (QP) charge of ${1\over4}e$ for the half-filled state
at $\nu_e={5\over2}$ \cite{Dolev08} and careful measurements 
of the minute excitation gaps \cite{Pan08,Choi08,Dean08}.
In theory, the most recent advances have been related to the breaking 
of particle-hole symmetry in the Moore--Read state \cite{Levin07}, 
the role of layer width \cite{Peterson08} and LL mixing 
\cite{LLmix,Simon08} in real systems, and the nature of QPs 
\cite{Toke07,Bernevig08}.

Despite intensive studies, connection of the FQH states in LL$_1$ 
($\nu_e={5\over2}$, ${7\over3}$, ${12\over5}$, or ${11\over5}$) 
to the few proposed wave functions is not conclusively established, 
and in some cases it is only tentatively assumed for the lack of 
other candidates.
This is an urgent problem, as the connection of some of these wave 
functions to the particular conformal field theories is precisely
what fuels the anticipation of non-Abelian statistics in nature.
Another unresolved puzzle is the discrepancy \cite{Dean08} between 
the experimental and numerical gaps at $\nu_e={5\over2}$, which  
undermines the understanding of this state in simple models assuming 
spin polarization or decoupling from the crystal lattice.
On the other hand, a wealth of IQLs found in various systems 
(electrons or CFs at different fillings of different LLs, in layers 
of varied width $w$) invites a more general question of possible IQLs 
with arbitrary interactions $V$.

In this Letter we demonstrate (numerically) that non-Abelian IQLs 
can only emerge in LL$_1$ in the narrow range of filling factors 
${7\over3}<\nu_e<{8\over3}$.
In this range, the known non-Abelian candidates at $\nu_e={5\over2}$ 
and ${12\over5}$ are closely examined (including finite layer width 
and LL mixing) and found to have favorable correlation energies 
and suggestive excitation gaps.
Outside of it, the Abelian ground states of noninteracting CFs 
carrying four flux quanta repeat in both lowest LLs.
The borderline $\nu={1\over3}$ ground state is adiabatically connected 
to the Laughlin state, but it has a smaller gap, distinct
short-range correlations, negative quasihole (QH) energy in narrow 
layers, and probably different quasielectrons (QEs).

We study spin-polarized $N$-particle systems on a unit sphere with
the magnetic monopole of strength $2Q(hc/e)$ inside \cite{Haldane83}.
In this geometry, LL$_n$ is a shell of angular momentum $\ell=Q+n$, 
and different $N$-body wave functions at the same $\nu$ are distinguished 
by a `shift' $\gamma$ between LL degeneracy and $\nu^{-1}N$.
In contrast to previous exact diagonalization calculations, we do not 
confine ourselves to the systems with particular $V$, but search for 
the universality classes $(\nu,\gamma)$ of gapped ground states with 
arbitrary interactions, attainable in realistic systems 
(of electrons or CFs) with the suitable choice of $n$ and $w$.

We begin by recalling that many-body dynamics in a degenerate LL 
is determined by an interaction pseudopotential, defined 
\cite{Haldane83} as the dependence of the pair energy $V$ on the 
relative angular momentum $m=1$, 3, \dots.
Moreover, $V_m$ induces particular correlations through its deviation 
from a straight line over any consecutive $m$'s \cite{five-half}.
Hence, it is known that low-energy spectra of $V_m$ are reproduced 
by a suitable effective pseudopotential $U_m$ with only a few 
(positive or negative) coefficients.

We have used $U=[U_1,U_3,U_5]$ in the search for IQLs with (nearly) 
arbitrary interactions.
Higher terms (causing long-range order) were ignored.
On the other hand, inclusion of $U_5$ was needed for an accurate
description of the two lowest (electron or CF) LLs known to
host IQL states, while it still allowed for useful graphical 
representation of the ground state properties in an effectively 
two-dimensional space of (normalized) parameters $U_m$.

\begin{figure}[t]
\includegraphics[width=3.4in]{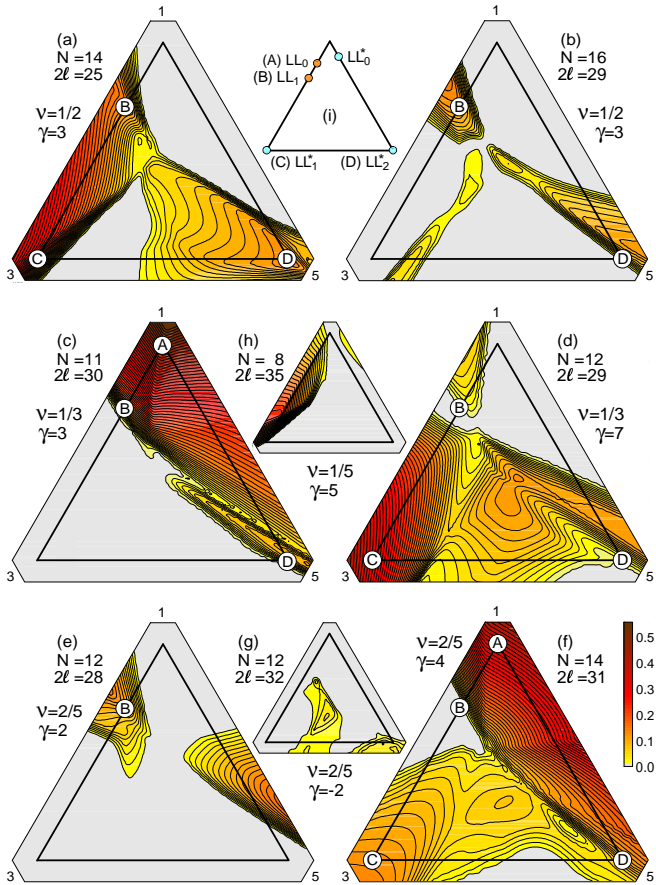}
\caption{(color online)
Ternary contour plots of the neutral excitation gaps $\Delta$
for $N$ fermions in a partially filled Landau level with shell 
angular momentum $\ell$, interacting via model pseudopotential 
$U_m$.
In each plot, three corners of the big triangles correspond to 
$U_m=\delta_{m,\mu}$ with $\mu=1$, 3, 5 marked in each corner.
Points relevant for the actual interactions in different electron 
or CF Landau levels are marked on a triangle in frame (i).
Different candidate incompressible states are indicated in frames 
(a)--(h), labeled by filling factors $\nu$ and shifts $\gamma$ 
assigned to each finite system $(N,2\ell)$.
For example, the Moore--Read state is marked as (B) in frames 
(a) and (b).}
\label{fig1}
\end{figure}

We looked at various finite systems $(N,2\ell)$ in search of 
the series of gapped ground states at $2\ell=\nu^{-1}N-\gamma$.
A few maps of the `neutral' gap $\Delta$ (gap to the first 
excited state in the same spectrum) appear in Fig.~\ref{fig1}.
The IQL candidates are the maxima in $\Delta$ repeating 
regularly for different $N$.
Their location on the map must be compared with the actual
pseudopotentials in different LLs, whose approximate positions 
(A)--(D) are indicated in frame (i).

Let us begin with a half-filled LL represented by $2\ell=2N-3$.
The maximum near $U_{(B)}=[{2\over3},{1\over3},0]$ found for 
each $N$ is relevant for the electrons in LL$_1$, whose actual 
Coulomb pseudopotential is almost linear between $m=1$ and 5.
The maximum coincides with a peak in similar maps (not shown) 
of the squared overlap of the ground state of $U$ with the 
exact Moore--Read state, and also with an essentially zero-level 
minimum of the triplet Haldane amplitude \cite{3body}.
This confirms quite definitively the earlier expectations 
that the Moore--Read ground state indeed occurs for a class of 
pseudopotentials close to that of LL$_1$ and that its accuracy 
depends sensitively on the fine-tuning of the leading $V_m$'s, 
achieved by adjusting the layer width $w$ \cite{Peterson08}.
Remarkably, it precludes the Moore--Read state in other 
half-filled LLs (e.g., in the second CF LL, called LL$^*_1$,
with dominant repulsion at $m=3$ \cite{Lee01}, where the observed 
FQH state \cite{Pan03} must be different).

In similar maps for $\nu={1\over3}$, significant gaps are found 
in different areas for the series of states with $\gamma=3$
(Laughlin) and 7.
The latter was proposed earlier for both LL$_1$ \cite{five-half} 
and LL$^*_1$ \cite{clusters} but, despite numerical evidence for 
pairing, its wave function remains unknown.
It is clear from Fig.~\ref{fig1}(c) and (d) that in the vicinity 
of $U_{(B)}$, relevant for LL$_1$, competition between the 
$\gamma=3$ and 7 states must be (and will be, in the following)
resolved more carefully.

For $\nu={2\over5}$, Jain's series with $\gamma=4$ correctly 
represents the ground state in LL$_0$.
In LL$_1$, this series competes with two others: the parafermion 
$\gamma=-2$ state \cite{Read99} and a recently proposed 
\cite{Bonderson08} state with $\gamma=2$.
Especially for the latter, Fig.~\ref{fig1}(e) appears suggestive 
of a gap emerging around $U_{(B)}$.
The subsequent careful analysis of the competition between these 
three universality classes is crucial, because the candidate 
states with $\gamma=\pm2$ are both non-Abelian, in contrast to 
the $\gamma=4$ Jain state.

Finally, at $\nu={1\over5}$, any positive $U=[U_1,U_3,0]$ yields 
an exact Laughlin state. 
Fig.~\ref{fig1}(h) shows that it is true description of the FQH 
states in both LL$_0$ and LL$_1$ \cite{Ambrumenil88}.
On the other hand, its relevance to the FQH effect observed in 
LL$^*_1$ \cite{Pan03} is doubtful (indeed, an alternative series 
with $\gamma=9$ appears to have more favorable correlations).

\begin{figure}
\includegraphics[width=3.4in]{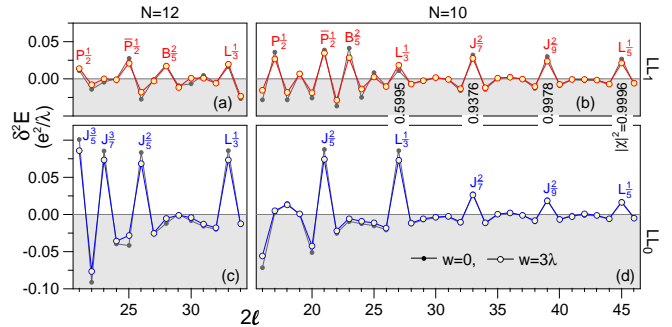}
\caption{(color online)
Symmetric second-order differences $\delta^2E$ of the ground-state 
energy per particle $E$ for $N=10$ and 12 electrons in LL$_0$ and 
LL$_1$, as a function of shell angular momentum $\ell$, for layer 
widths $w=0$ and $3\lambda$.
Candidate incompressible states are marked as `$\Phi\nu$', 
where $\Phi={\rm P}$, $\overline{\rm P}$, L, J, B denote 
Moore--Read pfaffian, anti-pfaffian, Laughlin and Jain states, 
and the state of Ref.~\cite{Bonderson08}.
Squared overlaps between the states repeating in both LLs at 
$\nu\le{1\over3}$ are indicated.}
\label{fig2}
\end{figure}

Guided by the maps of $\Delta$, pair/triplet amplitudes, 
and various overlaps we now focus on the FQH states in LL$_1$.
In Fig.~\ref{fig2} we seek confirmation of the IQL candidates 
in cusps of the dependence of the ground-state energy per 
particle $E$ on the LL degeneracy.
The cusps are most pronounced in the plots of $\delta^2E_{2\ell}
=E_{2\ell-1}+E_{2\ell+1}-2E_{2\ell}$.
For an IQL, its (positive) value gives the QP gap $\tilde\Delta$ 
times the number of QPs created per flux quantum.

Besides complementing the maps of Fig.~\ref{fig1} in the 
identification of the universality classes of particular IQLs, 
Fig.~\ref{fig2} reveals a simple connection between statistics 
and filling factor in LL$_1$.
At $\nu<{1\over3}$ the same Laughlin/Jain IQLs occur in 
LL$_1$ and LL$_0$, with high overlaps and similar gaps.
This similarity, earlier pointed out in Ref.~\cite{Ambrumenil88}, 
is caused by sufficiently high $V_1$ and similar $V_{m\ge3}$ in 
both LLs.
It validates the noninteracting CF model \cite{Jain89} with 
four flux quanta attached to each electron in LL$_1$.

In contrast, at $\nu>{1\over3}$ the Jain sequence of LL$_0$ is
replaced in LL$_1$ by a different set of IQLs, including the 
pfaffian and anti-pfaffian pair at $\nu={1\over2}$ and, apparently, 
a different $\nu={2\over5}$ state with $\gamma=2$ (Fig.~\ref{fig2} 
shows no signatures of the parafermion $\nu={2\over5}$ series with 
$\gamma=-2$ \cite{Read99}).
The breakdown of the noninteracting two-flux CF model in LL$_1$
opens, exclusively at ${1\over3}<\nu<{2\over3}$, a possibility 
for other IQLs, including several suggested more exotic states 
with various non-Abelian statistics of their QPs.

For the borderline $\nu={1\over3}$ state, separating the Abelian 
from (possibly) non-Abelian states, our calculation reinforces 
the theory of Ref.~\cite{Ambrumenil88}.
This state has a moderate overlap with the Laughlin ground state 
of LL$_0$, despite falling into the same class of $\gamma=3$.
It also has a smaller gap than the other Laughlin/Jain states 
in LL$_0$ or LL$_1$.

\begin{figure}
\includegraphics[width=3.4in]{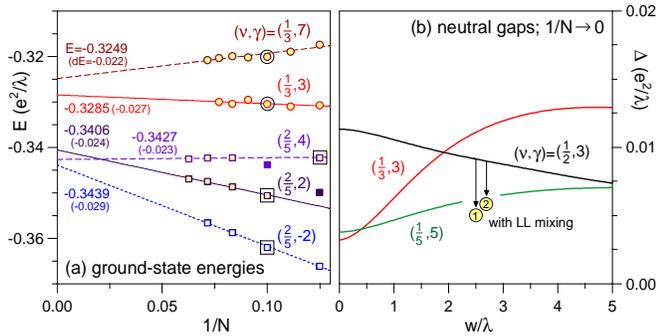}
\caption{(color online)
(a) Size extrapolation of the ground-state energies per particle 
$E$ calculated for $N$ electrons in the second Landau level.
Competing series of candidate incompressible states with filling 
factors $\nu={1\over3}$ and ${2\over5}$ are distinguished by their 
shifts $\gamma$.
Energy corrections $dE$ due to Landau level mixing were obtained
for $B=2.6$~T and $N=8$ or 10, as indicated with large open symbols.
(b) Dependence of the extrapolated neutral gaps $\Delta$ on the 
layer width $w$, for the $\nu={1\over2}$, ${1\over3}$, and 
${1\over5}$ nondegenerate ground states in the second Landau level.
For $\nu={1\over2}$, circles (1) and (2) show gap reduction due 
to Landau level mixing relevant for the experimental Refs.~\cite{Dean08} 
and \cite{Pan08}, respectively.}
\label{fig3}
\end{figure}

Having narrowed the search for non-Abelian IQLs in LL$_1$ to the 
range of ${1\over3}<\nu<{2\over3}$, let us now decide more conclusively 
between the competing candidate wave functions for the particular 
filling factors.
For $\nu={1\over2}$, the issues of the interplay of pfaffian 
and anti-pfaffian states \cite{Levin07} and of spin polarization 
\cite{Morf98} were already discussed.
In Fig.~\ref{fig3}(a) we compare the ground-state energies per particle 
$E$ of three known candidates for $\nu={2\over5}$ (we used $N\le16$ 
and $w=0$; results for $w>0$ are similar; energies $E$ are rescaled by 
$\sqrt{2Q\nu/N}$ to ensure equal units $e^2/\lambda$ for each $\gamma$).
In the extrapolation we only used the open squares, discarding one 
state aliased with the anti-pfaffian and one apparently suffering 
from the small size.

Although Fig.~\ref{fig3}(a) suggests that the ground state at $\nu=
{2\over5}$ might have $\gamma=-2$ (despite the lack of significant 
gaps in Figs.~\ref{fig1} and \ref{fig2}), the extrapolated energies 
of all three competing states are indecisively close.
Hence, comparison of their susceptibility to LL mixing is crucial. 
We estimated the appropriate energy correction $dE$ by including in 
the diagonalization additional states involving a single cyclotron 
excitation \cite{LLmix}.
For each $\gamma$, the value of $dE$ was calculated for $N\le10$
(it was also checked that $dE$ is less $N$-dependent than $E$).
The assumed Coulomb-to-cyclotron energy ratio $\beta\equiv(e^2/\lambda)
/(\hbar\omega_c)=1.56$ corresponds to $B=2.6$~T.
Clearly, in addition to having the lowest $E$, the $\gamma=-2$ parafermion 
state also has the largest $|dE|$, and therefore it is convincingly 
predicted to define the universality class of the $\nu={2\over5}$ ground 
state in LL$_1$.
This is important because this state is the only known candidate IQL 
\cite{Nayak08} whose braiding rules are sufficiently complex to allow 
quantum computation.

Fig.~\ref{fig3}(a) presents also the energies for $\nu={1\over3}$.
Here, the $\gamma=7$ series is firmly ruled out in favor of the 
$\gamma=3$ states (adiabatically connected to the Laughlin state).

In Fig.~\ref{fig3}(b) we plot the extrapolated neutral gaps $\Delta(w)$ 
for several IQLs in LL$_1$.
Unlike in LL$_0$ and in agreement with the experiments \cite{Pan08,Choi08}, 
the gaps at $\nu={1\over2}$, ${1\over3}$, ${1\over5}$ are all similar, 
and the gap at $\nu={2\over5}$ is much smaller (not shown).
Understanding that $\Delta$ includes the QE--QH attraction, it is 
nevertheless noteworthy that incorporating an earlier estimate of 
gap reduction due to LL mixing at $\nu={1\over2}$ \cite{LLmix} gives 
a surprisingly good agreement with experiments in the high mobility 
limit:
for Refs.~\cite{Pan08,Choi08}, $w/\lambda=2.7$ and $\beta=1.1$, 
yielding the 35\% reduction and $\Delta=0.006\,e^2/\lambda$;
for Ref.~\cite{Dean08}, $w/\lambda=2.5$ and $\beta=1.7$, 
yielding the 45\% reduction and $\Delta=0.005\,e^2/\lambda$.

Let us now examine more closely the crossover $\nu={1\over3}$ ground 
states in LL$_1$.
For $N\le12$, their squared overlaps with the Laughlin state are 
merely $\sim50\%$.
This reflects quite different correlations (e.g., for $N=12$ 
and $w=3\lambda$, the pair amplitudes at $m=1$, 3, 5, 7 are 
0.02, 0.13, 0.17, 0.10 in LL$_1$, as opposed to 0.00, 0.18, 
0.13, 0.09 in LL$_0$).
The magneto-roton band is also absent in the spectra of LL$_1$.
The low-energy states resembling Laughlin QEs and QHs are found 
at $2\ell=3N-3\mp1$, but they are not generally the lowest states 
in their spectra.
For example, the ground state at $2\ell=3N-4$ has angular 
momentum $L=2$, ${3\over2}$, 1, ${1\over2}$, 0 for $N=8$, 
\dots, 12 (instead of $L={1\over2}N$).

\begin{figure}
\includegraphics[width=3.4in]{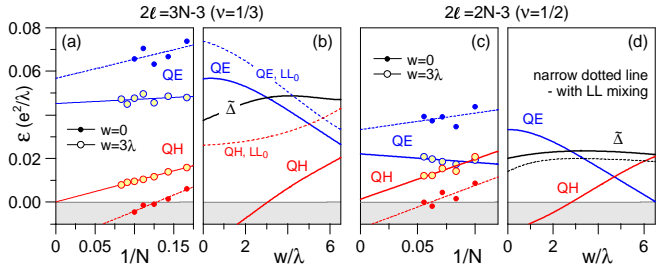}
\caption{(color online)
Quasiparticle energies $\varepsilon$ at the filling factors 
$\nu={1\over3}$ and ${1\over2}$ in the second Landau level:
(a,c) 
Size extrapolation of the $N$-electron results for the layer 
widths $w=0$ and $3\lambda$.
(b,d)
Extrapolated quasielectron and quasihole energies, and 
their sum $\tilde\Delta$, as a function of width $w$.
In (b), curves for the lowest Landau level are added for 
comparison.
In (d), dotted line includes Landau level mixing 
($B=2.6$~T).}
\label{fig4}
\end{figure}

Although we have checked that the eigenstates adiabatically connected 
to the Laughlin QPs become the lowest states under bias, their joint 
energy $\tilde\Delta=\varepsilon_{\rm QE}+\varepsilon_{\rm QH}$ is 
unreasonably high.
Fig.~\ref{fig4}(a,b) shows the size extrapolation of QE and QH 
energies $\varepsilon$, and width dependences of the extrapolated 
$\varepsilon$'s.
Remarkably, $\varepsilon_{\rm QH}>0$ requires finite width 
$w>3\lambda$.
Negative QH energies in narrow layers are also found in 
other states in LL$_1$ identified in Fig.~\ref{fig2}(b).
For instance, $\varepsilon={1\over2}(E_{2\ell\pm1}-E_{2\ell})$ 
plus the size corrections \cite{Morf02} are plotted
in Fig.~\ref{fig4}(c,d) for $\nu={1\over2}$.

Note that the gap $\tilde\Delta$ at $\nu={1\over2}$ is not 
immediately weakened in wider wells (in contrast to 
Ref.~\cite{Morf02} where only data for $N=10$ and 14 was 
used for $w>0$).
The initial gap enhancement in wider wells is even more 
pronounced with the inclusion of LL mixing (here, the 
corrections $dE$ were calculated for $N\le14$ at $2\ell
=2N-3$ and $2N-3\pm1$).
While the reasonable match between the experimental gaps 
and $\Delta$ (rather than $\tilde\Delta$) at $\nu={1\over2}$ 
remains a mystery, the difference between the latter two is 
well understood as the QE--QH atraction \cite{Morf02}.
However, the difference between $\Delta$ and $\tilde\Delta$ 
at $\nu={1\over3}$ is too large to be explained in this way.
Instead, Laughlin QEs are unlikely to be the elementary 
negative carriers in LL$_1$.

In conclusion, we have found the range of filling factors 
${1\over3}<\nu<{2\over3}$ in LL$_1$ in which 
the emergence of non-Abelian statistics is possible.
Inside this range, our calculations including finite layer 
width and LL mixing demonstrate that the spin-polarized ground 
states at $\nu={1\over2}$ and ${2\over5}$ are described by 
the ``pfaffian'' and ``parafermion'' wave functions (both 
non-Abelian).
Outside, the Jain states of noninteracting CFs repeat 
in both lowest LLs, precluding more exotic phases.
The borderline $\nu={1\over3}$ state is adiabatically connected 
to the Laughlin liquid but has a smaller gap and distinct 
excitations.

The author thanks Gunnar M\"oller, Nigel Cooper, and Guillaume Gervais
for many insightful comments, and acknowledges support from EU under 
the Marie Curie Intra-European Grant No.\ PIEF-GA-2008-221701 and from
the Polish MNiSW under grant N202-071-32/1513.

\end{document}